# A Novel Geometric Solution for Moving Target Localization through Multistatic Sensing in the ISAC System

Shun Zhuge
*School of Electrical and Electronic Engineering*
*NTU*
Singapore
szhuge002@e.ntu.edu.sg

Yugang Ma
*Institute for Infocomm Reasearch*
*A*STAR*
Singapore
mayg@i2r.a-star.edu.sg

Zhiping Lin
*School of Electrical and Electronic Engineering*
*NTU*
Singapore
ezplin@ntu.edu.sg

Yonghong Zeng
*Institute for Infocomm Reasearch*
*A*STAR*
Singapore
yhzeng@i2r.a-star.edu.sg

*Abstract*—This paper proposes a novel geometric solution for tracking a moving target through multistatic sensing. In contrast to existing two-step weighted least square (2SWLS) methods which use the bistatic range (BR) and bistatic range rate (BRR) measurements, the proposed method incorporates an additional direction of arrival (DOA) measurement of the target obtained from a communication receiver in an integrated sensing and communication (ISAC) system. Unlike the existing 2SWLS methods that require at least three transmitter-receiver (TX-RX) pairs to operate, the proposed algorithm can conduct location estimation with a single TX-RX pair and velocity estimation with two TX-RX pairs. Simulations reveal that the proposed method exhibits superior performance compared to existing 2SWLS methods, particularly when dealing with moderate levels of noise in DOA measurements.

## I. Introduction

Integrated sensing and communication (ISAC) stand out as a crucial enabling technology for forthcoming wireless communication standards, including B5G and 6G [1]-[5]. The defining feature of ISAC lies in its ability to employ a single ISAC waveform for both communicating with user equipment (UE) and simultaneously sensing the target [5].

Recent advancements in communication technology, particularly in 5G, have ushered in the use of broader signal bandwidths and massive multiple-input and multiple-output (mMIMO) techniques, enabling accurate estimation of range, speed, and angle with comparable performances to dedicated radar [6]-[11]. Thus, it is possible to use a single waveform based on the 5G New Radio (NR) standard to achieve both communication and sensing functions. In our earlier research, as outlined in [11], we introduced a multistatic ISAC system built upon the standardized 5G NR waveform. The transmitted signal contains modulated data, which allows the communication receiver to receive useful communication data while performing the sensing function simultaneously. The sensing function can detect bistatic radar parameters, including the bistatic range (BR), bistatic range rate (BRR), and direction of arrival (DOA) of the target. However, our previous work in [11] only discussed the estimation of those radar parameters, and the localization algorithm was not addressed. Therefore, this paper focuses on investigating the multistatic sensing algorithm to locate a moving target in the ISAC system.

Multistatic sensing involves multiple transmitter-receiver (TX-RX) pairs in a system and is an area of significant interest among researchers [6], [12]-[16]. Many existing localization algorithms are inspired by the well-known two-step weighted least square (2SWLS) method in [13]. Reference [14] proposes a 2SWLS-based closed-form solution to locate the target position with BR measurements. Reference [15] introduces a 2SWLS method for both location and velocity estimations. In [16], an improved 2SWLS-based closed-from method is proposed and outperforms methods presented in [12] and [15]. However, all those methods require a minimum of three TX-RX pairs for operation which may affect their practicality when available pairs are limited in practice. Moreover, the extra DOA measurement obtained in the ISAC system may improve the localization performance, but it is not used as an intermediate measurement in those methods mentioned above.

Motivated by those limitations, we propose a novel localization algorithm in an ISAC system through multistatic sensing in this paper. Our method utilizes BR, BRR, and DOA measurements obtained from our previous work in [11]. The contributions of this paper are summarized below:

- A novel representation of the geometric relationships among transmitters, the target, and the receiver. As BR and DOA measurements are known at the receiver, the location of the unknown target can be expressed in terms of all known parameters.

- A new multi-static sensing algorithm that requires only one TX-RX pair to estimate target position and only two TX-RX pairs to estimate target velocity. In contrast, the existing 2SWLS methods must use at least three TX-RX pairs to locate a target.

Compared with the existing localization method in [16] under the same number of TX-RX pairs, our proposed method can achieve lower root-mean-square error (RMSE) in both position and velocity estimation at a moderate estimation error of DOA.

The remainder of this paper is organized as follows. In Section II, we formulate the multistatic sensing problem of a moving target in an ISAC system using BR, BRR, and DOA measurements. The proposed algorithm is presented in Section III. Section IV gives the simulation results and compares the performance between our proposed method and the method in [16]. Finally, some conclusions are drawn in Section V.

## II. Measurement model

In this section, the measurement model is presented. As we consider an additional DOA measurement, the model is

This research is supported by the National Research Foundation, Singapore and Infocomm Media Development Authority under its Future Communications Research Development Programme (Grant award number: FCP-ASTAR-TG-2022-003).

built in polar coordinates. We demonstrate the derivation of how to represent the range and range rate in our proposed model. Noise is also considered in our measurement model.

*A. Problem Formation*

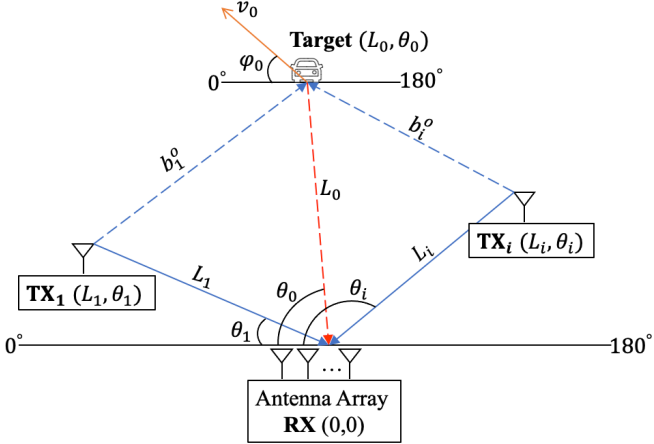

Fig. 1. The design of multistatic sensing in an ISAC system.

In this paper, we consider a 2-D multistatic sensing problem focused on tracking a moving target in an ISAC system, as depicted in Fig. 1. The communication transmitter (TX) sends a modulated signal based on the 5G NR standard to the ISAC communication receiver (RX). At the same time, the unknown target also reflects the signal. The RX with an antenna array can receive signals from both direct path and reflected path and function as a radar sensor for detecting the BR, BRR, and DOA associated with an unknown moving target [11]. The system model consists of $N$ TXs and one RX with known locations. Each TX, denoted as the $TX_i$ (with $1 \leq i \leq N$), is positioned at polar coordinates $(L_i, \theta_i)$, while the RX is located at the origin (0, 0). The unknown moving target is positioned at polar coordinates $(L_0, \theta_0)$ and in motion at a velocity $(v_0, \varphi_0)$.

*B. Range Representation*

We name the paths from the $TX_i$ to the RX and the target as the $i$-th direct path and the $i$-th target path respectively. The path from the target to the RX is called the reflected path. Given the polar coordinates of the $TX_i$, the RX, and the target, the ranges of the $i$-th direct path and $i$-th target path are denoted as $L_i$ and

$$b_i^o = \sqrt{L_i^2 - 2L_iL_0\cos(\theta_i - \theta_0) + L_0^2}, \quad (1)$$

respectively. The range of the reflected path is $L_0$. Thus, the BR with respect to the $i$-th TX-RX pair is obtained as

$$\Delta_i^o = L_0 + b_i^o - L_i. \quad (2)$$

*C. Range Rate Representation*

Denoting the rate of change of $L_0$ and $b_i^o$ as $\dot{L}_0$ and $\dot{b}_i^o$ respectively. As the target is the sole moving object in the system, $\dot{L}_0$ signifies the component of the target's velocity projected onto the direction of the reflected path. Similarly, $\dot{b}_i^o$ represents the component of the target's velocity projected onto the direction of the $i$-th target path. Given the varying TXs locations, we encounter two distinct cases.

*1) Case 1*: As illustrated in Fig. 2, the DOA of the $TX_i$ is smaller than the DOA of the target ($\theta_0 > \theta_i$). $\angle 1, \angle 2$ and $\angle 3$ are drawn in Fig. 2 for better presentation. By using the law of sines, the angle between the $i$-th target path and direct path can be expressed as

$$\alpha_i = \sin^{-1}\frac{L_0 \sin(|\theta_0 - \theta_i|)}{b_i^o}, \quad (3)$$

By using the alternate interior angles theorem, in Fig. 2,
$$\angle 1 = \theta_0. \quad (4)$$
By using the corresponding angles theorem,
$$\angle 2 = \angle 3. \quad (5)$$
By using the Euclidean exterior angle theorem,
$$\alpha_i = \angle 2 + \theta_i. \quad (6)$$
By substituting (5) into (6),
$$\angle 3 = \alpha_i - \theta_i. \quad (7)$$

Thus, from (4) to (7), by taking the target location as the origin, the directions of $\dot{L}_0$ and $\dot{b}_i^o$ are denoted as
$$\varphi_1 = \pi + \theta_0, \quad (8)$$
and
$$\varphi_2 = \pi - \alpha_i + \theta_i. \quad (9)$$

Therefore, by projecting the target velocity onto the direction of the reflected path and the direction of the $i$-th target path, $\dot{L}_0$ and $\dot{b}_i^o$ can be expressed as
$$\dot{L}_0 = v_0 \cos(\varphi_0 - \varphi_1) = v_0 \cos(\varphi_0 - \pi - \theta_0), \quad (10)$$
and
$$\dot{b}_i^o = v_0 \cos(\varphi_0 - \varphi_2) = v_0 \cos(\varphi_0 - \pi + \alpha_i - \theta_i). \quad (11)$$

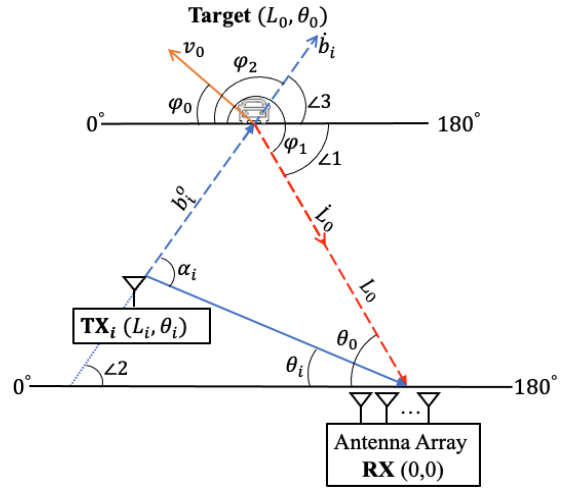

Fig. 2. Velocity projection when $\theta_0 > \theta_i$.

*2) Case 2*: As illustrated in Fig. 3, the DOA of the $TX_i$ is greater than the DOA of the target ($\theta_0 < \theta_i$). Similarly, $\angle 1, \angle 2$ and $\angle 3$ are drawn in Fig. 3 for better presentation. By using the law of sines, the angle between the $i$-th target path and direct path can be expressed as

$$\alpha_i = \sin^{-1}\frac{L_0 \sin(|\theta_0 - \theta_i|)}{b_i^o}, \quad (12)$$

By using the alternate interior angles theorem, in Fig. 3,
$$\angle 1 = \theta_0. \quad (13)$$
By using the corresponding angles theorem,
$$\angle 2 = \varphi_2. \quad (14)$$
By using the Euclidean exterior angle theorem,
$$\alpha_i = \angle 2 + \pi - \theta_i. \quad (15)$$

Thus, from (13) to (15), by taking the target location as the origin, the directions of $\dot{L}_0$ and $\dot{b}_i^o$ are denoted as
$$\varphi_1 = \pi + \angle 1 = \pi + \theta_0, \quad (16)$$
and
$$\varphi_2 = \angle 2 = \alpha_i + \theta_i - \pi. \quad (17)$$
Therefore, by projecting the target velocity onto the direction of the reflected path and the direction of the $i$-th target path, $\dot{L}_0$ and $\dot{b}_i^o$ can be expressed as,
$$\dot{L}_0 = v_0 \cos(\varphi_0 - \varphi_1) = v_0 \cos(\varphi_0 - \pi - \theta_0), \quad (18)$$
and
$$\dot{b}_i^o = v_0 \cos(\varphi_0 - \varphi_2) = v_0 \cos(\varphi_0 - \alpha_i - \theta_i + \pi). \quad (19)$$

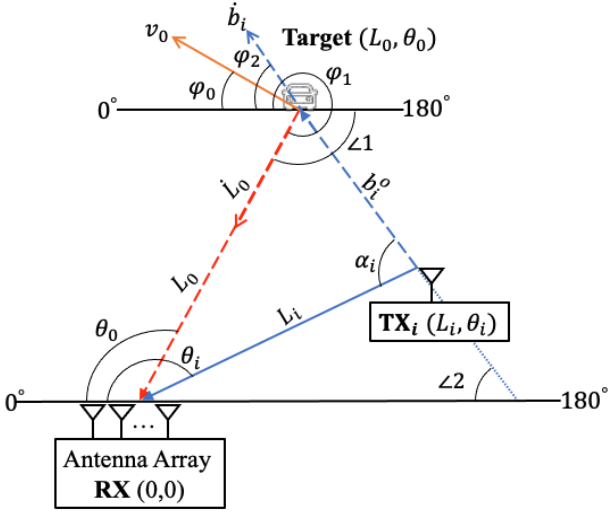

Fig. 3. Velocity projection when $\theta_0 < \theta_i$.

*3) General Case*: By comparing (10) with (18) and (11) with (19), we can express $\dot{L}_0$ and $\dot{b}_i^o$ in a general case without considering the location of the $TX_i$ as
$$\dot{L}_0 = v_0 \cos(\varphi_0 - \pi - \theta_0), \quad (20)$$
and
$$\dot{b}_i^o = v_0 \cos\bigl(\varphi_0 - \theta_i + \text{sign}(\theta_0 - \theta_i)(\alpha_i - \pi)\bigr), \quad (21)$$
where $\alpha_i = \sin^{-1}\frac{L_0 \sin(|\theta_0 - \theta_i|)}{b_i^o}$. Thus, from (20) and (21), the BRR with respect to $i$-th TX-RX pair is obtained as
$$\dot{\Delta}_i^o = -\dot{L}_0 + \dot{b}_i^o. \quad (22)$$

*D. Noise Formulation*

Considering measurement noises in practice, the BR, BRR and DOA measurements with respect to the $i$-th TX-RX pair are obtained as
$$\Delta_i = \Delta_i^o + n_i, \quad (23)$$
$$\dot{\Delta}_i = \dot{\Delta}_i^o + \dot{n}_i, \quad (24)$$
and
$$\beta_i = \theta_0 + \hat{n}_i, \quad (25)$$
where $n_i$, $\dot{n}_i$ and $\hat{n}_i$ are additive zero-mean Gaussian noises with variances, $\sigma_1^2$, $\sigma_2^2$, and $\sigma_3^2$ respectively. Obviously, for total $N$ TX-RX pairs in the system, there are $N$ BR measurements, $N$ BRR measurements and $N$ DOA measurements. The objective of this paper is to determine the position and velocity of the target from those noisy BR, BRR, and DOA measurements obtained in the ISAC system.

### III. LOCALIZATION ALGORITHM

In this section, a novel localization algorithm based on the BR, BRR, and DOA measurements is presented. Denote the estimated target position and velocity as $(\hat{L}_0, \hat{\theta}_0)$ and $(\hat{v}_0, \hat{\varphi}_0)$.

*A. Position Estimation*

By substituting (1) into (2), we can re-express the BR for the $i$-th TX-RX pair as
$$\Delta_i^o = L_0 + \sqrt{L_i^2 - 2L_i L_0 \cos(\theta_i - \theta_0) + L_0^2} - L_i. \quad (26)$$
Rearranging (26) and squaring both sides of it, we can obtain the following equation for this single TX-RX pair,
$$(\Delta_i^o - L_0 + L_i)^2 = L_i^2 - 2L_i L_0 \cos(\theta_i - \theta_0) + L_0^2. \quad (27)$$
From (27), we can express $L_0$ in terms of $\Delta_i^o$, $L_i$, $\theta_i$, and $\theta_0$ as
$$L_0 = \frac{\Delta_i^{o2} + 2\Delta_i^o L_i}{2\Delta_i^o L_i + 2L_i - 2L_i \cos(\theta_i - \theta_0)}. \quad (28)$$
The noisy measurement $\beta_i$ can be taken as the initial estimation of $\theta_0$. By replacing $\Delta_i^o$ and $\theta_0$ with noisy measurements $\Delta_i$ and $\beta_i$, an estimated $L_0$ can be obtained as
$$\hat{L}_0 = \frac{\Delta_i^2 + 2\Delta_i L_i}{2\Delta_i L_i + 2L_i - 2L_i \cos(\theta_i - \beta_i)}. \quad (29)$$
As all variables on the right side of (29) are known, we can estimate the position of the unknown target.

For $N$ distributed TXs and 1 RX, $N$ TX-RX pairs are formed. For $1 \leq i \leq N$, the collection of $\beta_i$ is given as
$$\boldsymbol{\theta} = [\beta_1, \beta_2, \dots, \beta_N]^T. \quad (30)$$
The estimation of $\theta_0$ can be expressed as
$$\hat{\theta}_0 = \arg\min_{x \in [0, \pi]} \|x\mathbf{1}_N - \boldsymbol{\theta}\|, \quad (31)$$
where $\mathbf{1}_N$ is the $N \times 1$ vector of all ones. By replacing $\hat{L}_0$ in (29) with $a_i$ for better presentation, and using the estimation result in (31), we obtained
$$a_i = \frac{\Delta_i^2 + 2\Delta_i L_i}{2\Delta_i L_i + 2L_i - 2L_i \cos(\theta_i - \hat{\theta}_0)}. \quad (32)$$
For $1 \leq i \leq N$, the collection of $a_i$ is given as
$$\mathbf{L} = [a_1, a_2, \dots, a_N]^T. \quad (33)$$
The estimation of $L_0$ can be expressed as
$$\hat{L}_0 = \arg\min_{y \in [0, A]} \|y\mathbf{1}_N - \mathbf{L}\|, \quad (34)$$
where $A$ is the maximum range can be determined by the system. We can solve (31) and (34) based on the maximum likelihood principle.

*B. Velocity Estimation*

By substituting (20) and (21) into (22), we can re-express the BRR for the $i$-th TX-RX pair as
$$\dot{\Delta}_i^o = -v_0 \cos(\varphi_0 - \pi - \theta_0) + v_0 \cos\bigl(\varphi_0 - \theta_i + \text{sign}(\theta_0 - \theta_i)(\alpha_i - \pi)\bigr), \quad (35)$$
where $\alpha_i = \sin^{-1}\frac{L_0 \sin(|\theta_0 - \theta_i|)}{b_i^o}$. By replacing $L_0$ and $\theta_0$ with $\hat{L}_0$ and $\hat{\theta}_0$ respectively in (1), the estimated range of the $i$-th target path can be obtained as
$$\hat{b}_i = \sqrt{L_i^2 - 2L_i \hat{L}_0 \cos(\theta_i - \hat{\theta}_0) + \hat{L}_0^2}. \quad (36)$$
Similarly, by replacing $L_0$, $\theta_0$, $b_i^o$, $v_0$, and $\varphi_0$ with $\hat{L}_0$, $\hat{\theta}_0$, $\hat{b}_i$, $v$, and $\varphi$ respectively in (35), we obtain the estimated BRR as

$$\hat{\dot{\Delta}}_i = -v\cos(\varphi - \pi - \hat{\theta}_0) + v\cos\left(\varphi - \theta_i + \text{sign}(\hat{\theta}_0 - \theta_i)(\hat{\alpha}_i - \pi)\right), \quad (37)$$

where $\hat{\alpha}_i = \sin^{-1}\frac{\hat{L}_0 \sin(|\hat{\theta}_0 - \theta_i|)}{\hat{b}_i}$, $v \in [0, V]$, $\varphi \in [0, \pi]$, and $V$ is the maximum speed can be determined by this system. For $1 \leq i \leq N$, the collections of the estimated BRR, $\hat{\dot{\Delta}}_i$, and the noisy BRR measurement, $\dot{\Delta}_i$, are given as

$$\hat{\dot{\boldsymbol{\Delta}}} = \left[\hat{\dot{\Delta}}_1, \hat{\dot{\Delta}}_2, \ldots, \hat{\dot{\Delta}}_N\right]^T, \quad (38)$$

and

$$\dot{\boldsymbol{\Delta}} = \left[\dot{\Delta}_1, \dot{\Delta}_2, \ldots, \dot{\Delta}_N\right]^T. \quad (39)$$

The estimation of $(v_0, \varphi_0)$ can be expressed as

$$(\hat{v}_0, \hat{\varphi}_0) = \arg\min_{(v,\ \varphi)} \left\|\hat{\dot{\boldsymbol{\Delta}}} - \dot{\boldsymbol{\Delta}}\right\|. \quad (40)$$

We can solve (40) based on the maximum likelihood principle.

## IV. SIMULATIONS

In this section, we conduct simulations of our proposed algorithm. We consider an ISAC system consisting of one static RX with an antenna array and three static TXs, all with the aim of estimating the position and velocity of an unknown moving target. The positions of these TXs and the RX can be found in Table I. To evaluate the performance of our proposed algorithm, we compare it with an existing localization method in [16]. For brevity, we will refer to this method as 2WLS.

TABLE I

POSITIONS OF THE UE AND ISAC RECEIVER

| System | Polar | Cartesian |
|---|---|---|
| RX | (0, 0) | (0, 0) |
| TX$_1$ | (50, 0) | (−50, 0) |
| TX$_2$ | (20, 45) | $(-10\sqrt{2}, 10\sqrt{2})$ |
| TX$_3$ | (25, 135) | $(12.5\sqrt{2}, 12.5\sqrt{2})$ |

The RMSE calculations for the estimated target position and velocity are given by

RMSE (position) =

$$\sqrt{\frac{\sum_{m=1}^{M}\left(\hat{L}_{0,m}^2 - 2L_0\hat{L}_{0,m}\cos(\hat{\theta}_{0,m} - \theta_0) + L_0^2\right)}{M}}; \quad (41)$$

RMSE (velocity) =

$$\sqrt{\frac{\sum_{m=1}^{M}\left(\hat{v}_{0,m}^2 - 2v_0\hat{v}_{0,m}\cos(\hat{\varphi}_{0,m} - \varphi_0) + v_0^2\right)}{M}}, \quad (42)$$

where $(\hat{L}_{0,m}, \hat{\theta}_{0,m})$ and $(\hat{v}_{0,m}, \hat{\varphi}_{0,m})$ are the estimated position and velocity of the unknown moving target at the $m$-th trial. We run $M = 5000$ independent trials for both 2WLS and the proposed method. As mentioned in Section II.D, zero-mean Gaussian noises are added to the BR, BRR, and DOA measurements in each trial. The noisy BR, BRR, and DOA measurements have noise levels of $\sigma_1$, $\sigma_2$, and $\sigma_3$ respectively. The units of BR, BRR, and DOA used in simulations are meter ($m$), meter per second ($m/s$), and degree (°) respectively. The same units are used in their noise levels. We position the target at (50, 90) with a velocity of (20, 90) in polar coordinates. Given that DOA measurements are not utilized in the 2WLS, we perform simulations of our proposed method by setting $\sigma_3 = 0°, 0.5°$ and $1°$ separately.

First, we examine the RMSEs of the various methods as $\sigma_1$ increases from $10^{-1}$ $m$ to $10^1$ $m$ while keeping $\sigma_2$ constant at $0.1$ $m/s$. The results are presented in Fig. 4 and Fig. 5. When $\sigma_3$ is set to $0°$, our proposed method consistently outperforms 2WLS in both position and velocity estimation. However, when $\sigma_3$ is set to $0.5°$ and $1°$, 2WLS exhibits superior performance for low values of $\sigma_1$ in both position and velocity estimation. Our proposed method demonstrates superior performance as $\sigma_1$ increases to high values. Notably, error floors emerge, leading to the convergence of RMSE performance at elevated $\sigma_1$ levels, independent of the $\sigma_3$ value. This phenomenon occurs due to the dominance of the constant $\sigma_3$ in estimation errors when $\sigma_1$ is low. As $\sigma_1$ becomes relatively high, estimation errors primarily originate from $\sigma_1$, diminishing the impact of $\sigma_3$.

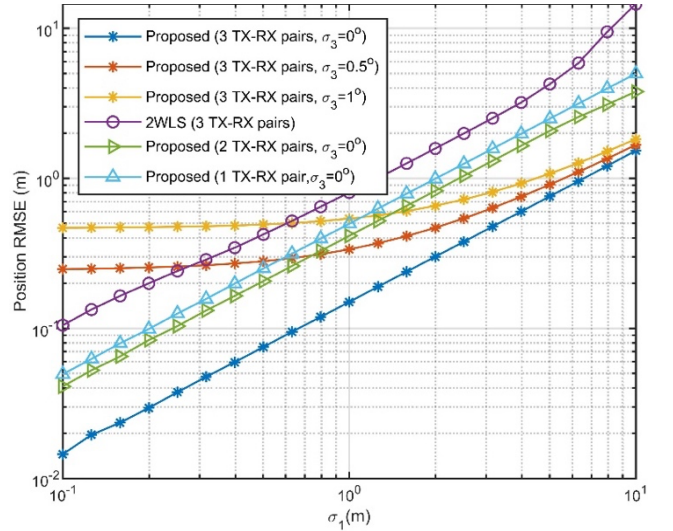

Fig. 4. Position RMSE comparison ($\sigma_2 = 0.1$ $m/s$).

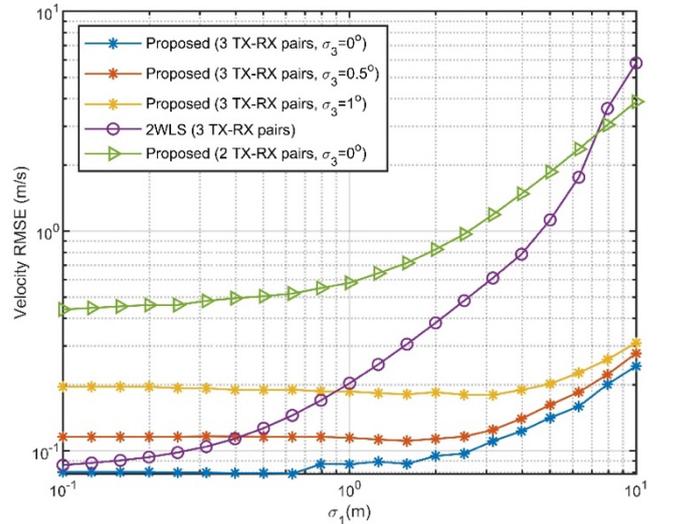

Fig. 5. Velocity RMSE comparison ($\sigma_2 = 0.1$ $m/s$).

Next, we examine the RMSEs of different methods as $\sigma_2$ increases from $10^{-1}$ $m/s$ to $10^1$ $m/s$ while $\sigma_1$ is fixed at $0.1$ $m$. The results are illustrated in Fig. 6 and Fig. 7. When $\sigma_3$ is set to $0°$, our proposed method consistently outperforms 2WLS in both position and velocity estimation. For position estimation, the RMSEs in Fig. 6 appear as horizontal lines across all scenarios. This is because BRR measurements have a relatively low impact on position estimation for both methods. When $\sigma_3$ is set to $0.5^o$ and $1^o$, 2WLS outperforms our proposed method in velocity estimation at very low $\sigma_2$ values, but our proposed method excels for higher $\sigma_2$ values. Error floors are again observed for our method in Fig. 7, where the RMSE performance converges to a similar level with varying $\sigma_3$ as $\sigma_2$ increases. This behavior occurs because at low $\sigma_2$, the constant values of $\sigma_1$ and $\sigma_3$ dominate velocity estimation errors. However, as $\sigma_2$ becomes relatively high, the errors are primarily influenced by $\sigma_2$, with the contributions of $\sigma_1$ and $\sigma_3$ diminishing in significance.

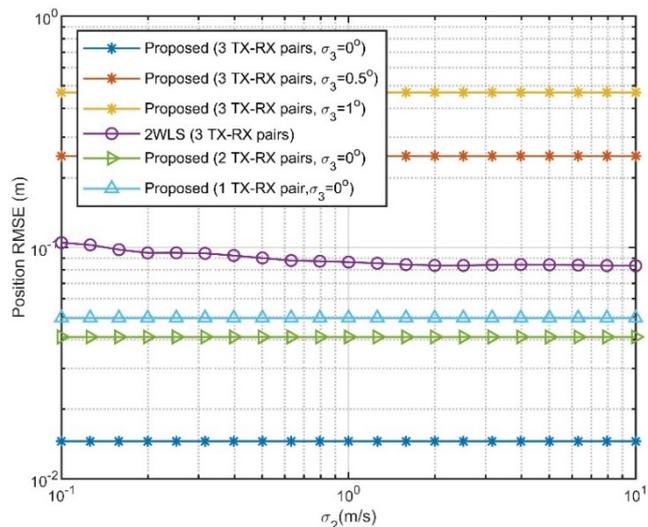

Fig. 6. Position RMSE comparison ($\sigma_1 = 0.1\ m$).

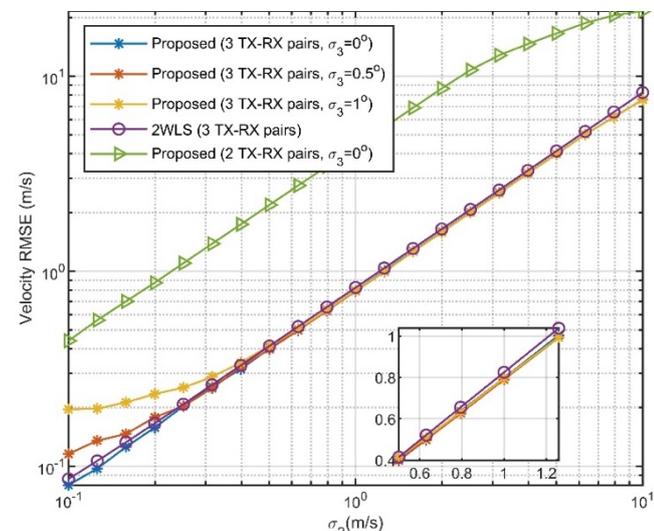

Fig. 7. Velocity RMSE comparison ($\sigma_1 = 0.1\ m$).

We also present the performance of our proposed method with 1 TX-RX pair and 2 TX-RX pairs in Fig. 4 to Fig. 7 under perfect DOA estimation. Fig. 4 and Fig. 6 demonstrate that our proposed method can outperform 2WLS in position estimation even when fewer TX-RX pairs are used.

## V. CONCLUSION

In conclusion, we introduce a novel localization algorithm tailored for tracking a moving target in an ISAC system, operating in a multistatic sensing scenario. Our method utilizes BR, BRR, and DOA measurements detected from the RX with an antenna array. A distinctive feature of our proposed method is its capability to estimate target position with just one TX-RX pair. To assess the effectiveness of our approach, we conduct a comparative analysis with an existing closed-form localization method that relies solely on BR and BRR measurements. Our findings demonstrate the superior performance of our proposed method in accurately estimating both the position and velocity of the target, particularly under conditions characterized by moderate levels of noise in DOA measurements.

As part of our future research, we will extend this work to more practical scenarios with multiple targets.